\documentstyle[prl,aps]{revtex} 
\def\Paris{Par\'{\i}s}
\def\Perez{P\'erez}
\def\Dominguez{Dom\'\i{}nguez}
\def\Astrofisica{Astrof\'{\i}sica}
\def\Fisica{F\'{\i}sica}
\def\Astrobiologia{Astrobiolog\'{\i}a}
\def\Torrejon{Torrej\'on}
\begin{document} 
\twocolumn
\title{ 
Effective potential for classical field theories subject to stochastic noise
} 
\author{ 
David Hochberg$^{+,*}$, Carmen Molina--\Paris$^{++}$,  
Juan \Perez--Mercader$^{+++,*}$, and Matt Visser$^{++++}$ 
} 
\address{
$^{+,+++}$Laboratorio de \Astrofisica\ Espacial y \Fisica\
Fundamental, Apartado 50727, 28080 Madrid, Spain\\
$^{++}$Theoretical Division, Los Alamos National Laboratory, Los
Alamos, New Mexico 87545, USA\\
$^{++++}$Physics Department, Washington University, Saint Louis,
Missouri 63130-4899, USA\\
$^{*}$Centro de \Astrobiologia, 
INTA, Ctra. Ajalvir, Km. 4, 28850 \Torrejon\  Madrid, Spain
}
\date{7 April 1999; 18 June 1999; 28 November 2000; \LaTeX-ed \today} 
\maketitle 

{\small 
Classical field theories coupled to stochastic noise provide an
extremely powerful tool for modeling phenomena as diverse as
turbulence, pattern-formation, and the structural development of the
universe itself. In this Letter we sketch a general formalism that
maps such systems into a field theory language, and demonstrate how to
extract the one-loop physics for an {\em arbitrary} classical field
theory coupled to Gaussian noise.  The {\em amplitude} of the noise
two-point function serves as the loop-counting parameter and is the
analog of Planck's constant $\hbar$. We define the effective action
and the effective potential, and derive a general formula for the
one-loop effective potential of a classical field theory coupled to
translation-invariant additive Gaussian noise. }


\bigskip
\noindent
V1: 7 April 1999.
 
\bigskip
\noindent
V2: 18 June 1999: Cosmetic changes to improve handling of
Fadeev--Popov ghosts.

\bigskip
\noindent
V3: 28 November 2000: References updated.
This paper has now been superseded and the contents of this paper have
now been incorporated into cond-mat/9904215 [Phys. Rev. {\bf E60} (1999)
6343-6360]; cond-mat/9904391 [Physica {\bf A280} (2000) 437-455]; and
cond-mat/9904413 [Phys. Lett. A, in press].

\newcommand{\Str}{\mathop{\mathrm{Str}}} 
\newcommand{\tr}{\mathop{\mathrm{tr}}} 
\newcommand{\define}{\mathop{\stackrel{\rm def}{=}}}
\newcommand{\Tr}{\mathop{\mathrm{Tr}}}
\def\d{{\mathrm{d}}}
\def\a2{a_{d/2}}
\def\implies{\Rightarrow}
\def\dirac{\gamma^\mu (\partial_\mu - A_\mu)}
\def\half{ {\scriptstyle{1\over2}} }
\def\A{ {\cal A} }
\def\PP{ {\cal P} }

Classical field theories subject to stochastic
noise~\cite{MSR,Zinn-Justin}, [e.g., stochastic partial differential
equations (SPDEs)], provide powerful means for modeling a host of
physical phenomena ranging from turbulence~\cite{Frisch}, to
pattern-formation~\cite{KPZ}, and to the structural development of the
universe itself~\cite{Berera-Fang,Hochberg-Mercader,PGHL,BDGP}.  In
this Letter we provide a brief summary of more extensive work to be
presented in~\cite{HMPV-spde}.  Consider the general class of SPDEs
\begin{equation}\label{spde}
D \phi(\vec x,t) = F[\phi(\vec x,t)] + \eta(\vec x,t).
\end{equation}
Here $D$ is any linear differential operator, involving arbitrary time
and space derivatives, which does {\em not} explicitly involve the
field $\phi$.  The function $F[\phi]$ is any forcing term, generally
nonlinear in the field $\phi$.  The forcing term will typically not
contain any time derivatives, but this is not an essential part of the
analysis.  The function $\eta(\vec x,t)$ denotes the source of
noise. As yet the nature and probability distribution of the noise
need not be specified.  We will take this SPDE and analyze it using
functional integral techniques.  In this Letter we sketch out the
field theory via the most direct route, avoiding the
Martin--Siggia--Rose construction (with its extra unphysical conjugate
fields used for book-keeping purposes~\cite{MSR,Zinn-Justin}).  We
will assume that the partial differential equation (\ref{spde}), plus
initial and boundary conditions, is well-posed. Given a
particular realization of the noise, $\eta$, the differential equation
is assumed to have a unique solution $\phi_{\mathrm soln}(\vec
x,t|\eta)$.  For any function $Q(\phi)$ of the field $\phi$ we can
define the stochastic average, (over the noise), as
\begin{equation}
\langle Q(\phi) \rangle \equiv
\int ({\cal D} \eta) \; \PP[\eta] \; Q(\phi_{\mathrm soln}(\vec x, t|\eta)).
\end{equation}
Here $\PP[\eta]$ is the probability density functional of the
noise. It is normalized to $1$, but is otherwise completely
arbitrary. We next use a functional delta-function to write the
following identity
\begin{eqnarray}
\phi_{\mathrm soln}(\vec x,t|\eta) 
&\equiv&
\int ({\cal D} \phi) \; \phi \;
\delta[\phi - \phi_{\mathrm soln}(\vec x,t|\eta)] 
\nonumber\\
&=&
\int ({\cal D} \phi) \; \phi \;
\delta[D\phi - F[\phi] - \eta] \;
\sqrt{ {\cal J} {\cal J}^\dagger }.
\end{eqnarray} 
The Jacobian functional determinant, ${\cal J}$, is defined by
\begin{equation}
{\cal J} \equiv \det\left( D - {\delta F\over\delta\phi} \right).
\end{equation}
The above is the functional analogue of a standard delta-function
result. If $f(x) =0$ has a unique solution at $x=x_0$, then
$x_0 = 
\int \d x \; x \; \delta(x-x_0) = 
\int \d x \; x \; \delta(f(x)) \; |f'(x)| 
=
\int \d x \; x \; \delta(f(x)) \; \sqrt{f'(x) [f'(x)]^*}.$
It is easy to see that one also has the identity
\begin{eqnarray}
&&
Q(\phi_{\mathrm soln}(\vec x,t|\eta)) 
\equiv
\nonumber\\
&&\qquad
\int ({\cal D} \phi) \; Q(\phi) \;
\delta [D\phi - F[\phi] - \eta] \;
\sqrt{{\cal J} {\cal J}^\dagger }.
\end{eqnarray} 
The ensemble average over the noise becomes
\begin{eqnarray}
\langle Q(\phi) \rangle &=&
\int ({\cal D} \eta) \; \PP[\eta] \times \; 
\nonumber\\
&& \int ({\cal D} \phi)\; Q(\phi) \;
\delta [D\phi - F[\phi] - \eta] \;
\sqrt{{\cal J} {\cal J}^\dagger }.
\end{eqnarray}
The noise integral is easy to perform, with the result that
\begin{equation}
\langle Q(\phi) \rangle =
\int  ({\cal D} \phi)\; \PP[D\phi-F[\phi]] \; Q(\phi) \;
\sqrt{{\cal J} {\cal J}^\dagger }.
\end{equation}
The effect of the noise appears in the stochastic average {\em only}
through its probability distribution $\PP[D\phi-F[\phi]]$.  The
presence of the functional determinant is essential; it is often
discarded without comment in the literature, but in general it must be
kept to ensure proper counting of the solutions to the original
SPDE. A particularly useful quantity is the characteristic functional.
With an obvious condensed notation, $\d x = \d^d\vec x \; \d t$, we
define
\begin{eqnarray}
Z[J] &\equiv& 
\left\langle \exp\left( \int J(x) \; \phi(x) \; \d x  \right) \right\rangle
\nonumber\\
&=& \int ({\cal D} \phi)\; \PP[D\phi - F[\phi]]
         \exp\left( \int J \; \phi \; \d x \right) \;
         \sqrt{{\cal J} {\cal J}^\dagger }.
\nonumber\\
\end{eqnarray}
When there is no risk of confusion we will suppress the $\d x$
completely.  This key result will enable us to calculate the effective
action and the effective potential in a direct way.  At this stage we
make some assumptions about the noise: We assume the noise is Gaussian
with zero mean (so that the only non-vanishing cumulant is the second
order one). If the noise has a non-zero mean, one can always redefine
the forcing term $F[\phi]$ to make the noise be of mean zero. We still
do not need to assume translation invariance, nor do we need to assume
the noise is white, power-law, or colored. Arbitrary Gaussian noise
is enough since it allows us to write the noise probability
distribution as
\begin{eqnarray}
\PP[\eta] &=& {1\over\sqrt{\det(2\pi G_\eta)}} 
\nonumber\\
&&
\exp\left[ -{1\over2} 
\int \int \d x \; \d y \; \eta(x) \;  G_\eta^{-1}(x,y) \; \eta(y) 
\right].
\end{eqnarray}
The characteristic function is thus
\begin{eqnarray}
\label{characteristic}
Z[J] &=& {1\over\sqrt{\det(2\pi G_\eta)}}
         \int ({\cal D} \phi)\; 
\sqrt{{\cal J} {\cal J}^\dagger }
         \exp\left( \int J \phi \right) \;
\nonumber\\
&&
         \exp\left[ -{1\over2}
                   \int \int (D\phi-F[\phi])  G_\eta^{-1} (D\phi-F[\phi])
             \right].
\nonumber\\
\end{eqnarray}
This characteristic function contains all the physics describable by
(\ref{spde}). The noise $\eta$ has been completely eliminated and
survives only through the explicit appearance of its two-point
function $ G_\eta$.  Since the characteristic function is now given as
a functional integral over the physical field $\phi$, all the standard
machinery of statistical and quantum field theory can be brought to
bear.  This formula for the characteristic function demonstrates that
(modulo Jacobian determinants) all of the physics of any stochastic
differential equation can be extracted from a functional integral
based on the ``classical action'' (note: this is {\it not} yet the
{\it effective} action)
\begin{equation}
{\cal S} [\phi] = \frac{1}{2}
  \int \int (D\phi-F[\phi])  G_\eta^{-1} (D\phi-F[\phi])
\; .
\end{equation}
This action $S[\phi]$ can be viewed as a generalization of the
Onsager--Machlup action~\cite{Onsager-Machlup}. Onsager and Machlup
dealt with stochastic mechanics rather than field theory, and limited
attention to white noise. As they developed their formalism with the
notions of linear response theory in mind, they tacitly assumed the
``forcing term'' $F[\phi]$ to be linear, so that {\em both} the noise
and the field fluctuations were Gaussian.

The Jacobian functional determinant is sometimes (but not always)
constant.  This is a consequence of the causal structure of the theory
as embodied in the fact that we are only interested in {\em retarded}
Green functions.  The situation here is in marked contrast to that in
QFT, where the relativistic nature of the theory forces the use of
{\em Feynman} Green functions ($+i\epsilon$
prescription)~\cite{HMPV-spde}.  In order to avoid enumerating special
cases, and to have a formalism that can handle both constant and
field-dependent Jacobian factors, we exponentiate the determinant via
the introduction of Fadeev--Popov ghost
fields~\cite{Zinn-Justin,HMPV-spde,Weinberg-1,Rivers}
\begin{equation}
{\cal J} =
{1\over\det{(2\pi I)}}
\int ({\cal D} [g,g^\dagger]) \;
\exp\left[
- \frac{1}{2} 
\int  g^\dagger \left( D - {\delta F\over\delta\phi} \right) g 
\right].
\end{equation}
The $g$ and $g^\dagger$ fields are known as scalar ghost fields. They
are defined in terms of anticommuting complex variables and behave in
a manner similar to an ordinary complex scalar field except that there
is an extra minus sign for each ghost loop. In all cases we are
interested in the operator $(D-\delta F/\delta\phi)$, even if not
self-adjoint, is real and non-singular. This allows us to replace
$\sqrt{{\cal J}^\dagger {\cal J}}$ by ${\cal J}$, so that the
characteristic function (\ref{characteristic}) can be written as
\begin{eqnarray}
Z[J] &=& 
{1\over\sqrt{\det[(2\pi)^3 G_\eta]}}
\int ({\cal D} [\phi,g,g^\dagger]) \;   
\exp\left( \int J \phi \right)
\nonumber\\
&&
\exp\left[ 
-{1\over2} \int \int (D\phi-F[\phi]) G_\eta^{-1}(D\phi-F[\phi])
\right] \;
\nonumber\\
&&
\exp\left[
-{1\over2}\int  g^\dagger \left( D - {\delta F\over\delta\phi} \right) g 
\right].
\end{eqnarray}
This procedure trades off the functional determinants for two extra
functional integrals. In order to develop a Feynman diagram expansion
we treat the driving term $F[\phi]$ as the perturbation and expand
around the free-field theory defined by setting $F=0$.  There are
two propagators, one for the $\phi$ field, and one for the
ghost fields.  For simplicity, we will now take the noise to be
translation invariant~\cite{HMPV-spde}. Translation-invariance lets us
take simple Fourier transforms in the difference variable $x-y$ (more
precisely: $\vec x-\vec y$ and $t_x-t_y$), so that in
momentum-frequency space we have
\begin{eqnarray}
G_{\mathrm field}(\vec k,\omega) &=& 
{ G_\eta(\vec k,\omega)\over D(-\vec k,-\omega) \; D(\vec k,\omega)}
\\
G_{\mathrm ghost}(\vec k,\omega) &=& {1\over D(\vec k,\omega)}.
\end{eqnarray}
The Feynman diagram vertices are formally~\cite{HMPV-spde}
\begin{eqnarray}
&\phi-F[\phi]\hbox{ vertex:}& \qquad
-{D(\vec k,\omega) \; \phi \; F[\phi]\over  G_\eta(\vec k,\omega)}
\\
&F[\phi]-F[\phi] \hbox{ vertex:}&  \qquad
+{1\over2}{ F[\phi] \; F[\phi] \over G_\eta(\vec k,\omega)}
\\
&\hbox{ghost vertex:}& \qquad
- \frac{1}{2} \; g^\dagger \; {\delta F[\phi]\over\delta\phi} \; g.
\end{eqnarray}
In practical calculations $F[\phi]$ is a polynomial in the field
and/or its gradients~\cite{HMPV-spde}.  Some specific examples are
discussed in~\cite{HMPV-kpz,HMPV-rdd}.

To set up the effective action corresponding to (\ref{spde}), and its
loop expansion, it is useful to separate the two-point function for
the noise into a {\em shape}, $g_2(x,y)$, and a constant {\em
amplitude}, $\A$
\begin{equation}
G_\eta(x,y) \define \A \; g_2(x,y).
\end{equation}
We adopt the following simplifying normalization of the shape function
$g_2(x,y)$
\begin{equation}
\int g_2^{-1}(\vec x,t)\; \d^d \vec x\; \d t 
\; = \; 1
\; = \; \tilde g_2^{-1}(\vec k=\vec 0,\omega=0).
\end{equation}
This is only a {\em convention}, not a restriction on the noise, since
it only serves to give an absolute meaning to the normalization of the
amplitude $\A$~\cite{HMPV-spde}.  The virtue of introducing the
amplitude parameter $\A$ is that it is the loop-counting parameter for
the field theory discussed in this Letter~\cite{HMPV-spde}. To verify
this, start by writing the characteristic function (with external
sources rescaled for convenience)
\begin{eqnarray}
Z[J] &=& 
{1\over\sqrt{\det[(2\pi)^3 G_\eta]}}
\int ({\cal D} [\phi,g,g^\dagger])\; 
\exp\left( {\int J \; \phi\over \A} \right)
\nonumber\\
&&
\exp\left[
-{1\over2} \int \int { (D\phi-F[\phi])  g_2^{-1} (D\phi-F[\phi]) \over\A }
\right] \;
\nonumber\\
&&
\exp\left[
-{1\over2}\int 
g^\dagger \left( D - {\delta F\over\delta\phi} \right) g 
\right].
\end{eqnarray}
The generating function for connected correlation functions is defined
by~\cite{Weinberg-1,Rivers}
\begin{equation}
\label{connected}
W[J] = + \A \left( \ln Z[J] - \ln Z[0] \right),
\end{equation}
and the effective action by~\cite{Weinberg-1,Rivers}
\begin{equation}
\Gamma[\phi;\phi_0] = - W[J] + \int \phi \; J; \; \; \; 
{\delta \Gamma[\phi;\phi_0]\over \delta \phi} = J,
\end{equation}
with $\phi_0$ the stochastic expectation value of $\phi$ for vanishing
external sources, $J=0$.  The previous equation defines the
non-perturbative effective action. For specific examples it is not
very useful, and we often have to restrict ourselves to a perturbative
calculation of the effective action, after singling out an expansion
parameter.  One can always develop a Feynman diagram expansion
provided that the classical action can be separated into a ``quadratic
piece'' (to define propagators) and an ``interacting term'' (to define
vertices).  In this expansion the sum of all connected diagrams
coupled to external sources $J(x)$ is exactly $W[J]$, and the
effective action $\Gamma[\phi;\phi_0]$ corresponds to all (amputated)
one-particle irreducible graphs (1PI), that is, Feynman diagrams that
cannot be made disconnected by cutting only one propagator.

To see that $\A$ is the loop-counting parameter, note that each field
propagator is proportional to $\A$ while each ghost propagator is
independent of $\A$. The vertices that do not include ghosts are
proportional to $\A^{-1}$, while ghost vertices are independent of
$\A$. But each ghost loop contains exactly as many ghost propagators
as ghost vertices.  Thus if one assigns a factor $\A$ to each ghost
propagator and a factor of $\A^{-1}$ to each ghost vertex, one will
not change the total number of factors of $\A$ assigned to the Feynman
diagram. Each diagram is proportional to $\A^{I-V}$, where $I$ is the
number of (internal) propagators in the diagram and $V$ is the number
of vertices.

It is a standard topological theorem that for any graph (not just any
Feynman diagram) $I - V = L - 1$, where $L$ is the number of
loops~\cite{Weinberg-1,Rivers}. This implies that stochastic field
theories based on SPDEs exhibit exactly the same loop-counting
properties as QFTs, except that the loop-counting parameter is now the
{\em amplitude} of the noise two-point function (instead of Planck's
constant $\hbar$).  The only subtle part of the argument has been in
dealing with the Fadeev--Popov ghosts, and it is important to realize
that this argument is completely independent of the details of the
differential operator $D$ and the forcing term $F[\phi]$. When it
comes to calculating the diagrams contributing to the effective
action, the extra explicit factor of $\A$ inserted in the definition
of $W[J]$, [equation (\ref{connected})], guarantees that the 1PI
graphs contribute to $\Gamma[\phi;\phi_0]$ with a weight that is
exactly $\A^L$, which means that $\A$ is a {\em bona fide} expansion
parameter.

For any field theory, quantum or stochastic, the characteristic
function can be written as
\begin{equation}
Z[J] = \int {\cal D}\phi 
\exp\left({ -S[\phi] + \int J\; \phi \over a}\right),
\end{equation}
and the one-loop {\it effective action} as
\begin{eqnarray}
\Gamma[\phi;\phi_0] &=& S[\phi] 
+ \half a \ln \det (S_2[\phi])
\nonumber\\
&&-
 \left( \phi \to \phi_0 \right)
+ O(a^2).
\end{eqnarray}
Here $S_2 \define \delta^2 S/\delta \phi(x) \delta \phi(y)$ is the
matrix of second functional derivatives of the action $S[\phi]$.  The
notation $S[\phi_0]$ is shorthand for $S[\langle\phi[J=0]\rangle]$.
When we consider field theories based on SPDEs, the loop-counting
parameter $a$ becomes $\A$, and the bare action is replaced
by~\cite{HMPV-spde}
\begin{eqnarray}
\label{action}
S[\phi] 
&\to&
S[\phi]
- \half\A \; \left(
\ln {\cal J} +
\ln {\cal J}^\dagger
\right).
\end{eqnarray}
The noise at this stage is translation-invariant and
Gaussian. Inserting equation (\ref{action}) into the formula for the
one-loop effective action, we obtain the general result (applicable to
all SPDEs)
\begin{eqnarray}
\Gamma[\phi;\phi_0] &=& S[\phi] 
\nonumber\\
&&+ \half{\A} \left\{ 
\ln \det (S_2[\phi]) - \ln {\cal J}[\phi] - \ln {\cal J}^\dagger[\phi] 
\right\}
\nonumber\\
&&
- \left( \phi \to \phi_0 \right)
+ O({\cal A}^2).
\end{eqnarray}
It is often useful to restrict to constant (spacetime independent)
fields. For such field configurations the effective action reduces to
the {\em effective potential} defined by
\begin{equation}
{\cal V}[\phi;\phi_0] \define {\Gamma[\phi;\phi_0]\over \Omega},
\end{equation}
where $\Omega=\int \d^d \vec x \, \d t$ is the volume of spacetime.
It is, in fact, sufficient to have both $D\phi=0$ and $F[\phi]={\rm
constant}$, the reason for this qualification will become clear when
we discuss the effective potential for the KPZ
equation~\cite{HMPV-kpz}. A straightforward
calculation~\cite{HMPV-spde} leads to
\begin{eqnarray}
\label{E-general}
{\cal V}[\phi] &=& 
\half F^2[\phi]
+\half \A \int {\d^d \vec k \; \d \omega\over (2\pi)^{d+1}} 
\nonumber\\
&&
\times
\ln  
\left[ 1 + {\tilde g_2{}(\vec k,\omega)
F[\phi] \; {\delta^2 F\over\delta\phi\;\delta\phi} \;
\over
\left( D^\dagger(\vec k,\omega)  - {\delta F\over \delta\phi}^\dagger \right)
\left( D(\vec k,\omega) - {\delta F\over \delta\phi} \right)}
\right]
\nonumber\\
&&-
 \left( \phi \to \phi_0 \right)
+ O(\A^2).
\end{eqnarray}
The above formula contains the most important result of this
Letter.  This result is qualitatively very reminiscent of the
effective potential for scalar QFT~\cite{Weinberg-1,Rivers}:
\begin{eqnarray}
{\cal V}[\phi] &=& 
V(\phi)
+ \half \hbar \int {\d^d \vec k \; \d \omega\over (2\pi)^{d+1}}
\ln  
\left[ 1 + { {\delta^2 V\over\delta\phi\;\delta\phi} 
\over
\omega^2 + \vec k^2 + m^2
}
\right]
\nonumber\\
&& -
 \left( \phi \to \phi_0 \right)
+ O(\hbar^2).
\end{eqnarray}
The major difference is the fact that the scalar QFT propagator is
replaced with a more complicated and typically nonrelativistic
propagator.  Notice also that for SPDEs one can naturally adapt the
noise to be both the source of fluctuations {\em and} the regulator to
keep the Feynman diagram expansion finite. This follows immediately by
inspection of (\ref{E-general}) which shows that the (momentum and
frequency dependent) noise shape function $\tilde g_2$ will affect the
momentum and frequency behaviour of the one-loop integral. The
finiteness, divergence structure, and renormalizability of this
integral will very much depend on the functional form of $\tilde g_2$.
Renormalizing such infinities on a case by case basis is relatively
simple since we are only working to one-loop~\cite{HMPV-kpz,HMPV-rdd}.

In this Letter we have shown how to set up a ``direct'' functional
formalism to explicitly calculate an effective action and a
corresponding effective potential for a general class of non-linear
SPDE's with additive noise. The effective action and effective
potential have the same structural definition as they do in quantum
field theory, though here they can be calculated even when the SPDE
does {\it not} follow from a Hamiltonian or Lagrangian variational
principle.  At the one-loop level in the noise amplitude we have
derived a compact expression for the effective potential reminiscent
of, but radically distinct from, the effective potential of scalar
QFT. The effective potential for stochastic phenomena provides a
powerful tool for exploring the influence of noise on a system, the
onset of pattern formation, and noise-induced dynamical symmetry
breaking.

We hope to have convinced the reader that the ``direct approach''
described in this Letter, and developed more fully
in~\cite{HMPV-spde}, is both useful and complementary to the more
traditional MSR formalism~\cite{MSR,Zinn-Justin}. Some questions can
more profitably be asked and answered in this ``direct''
formalism. For instance, our general formula for the one-loop
effective potential for {\em arbitrary} SPDEs subject to
translation-invariant Gaussian noise appears difficult to extract from
the MSR formalism.



\end{document}